\documentclass[twocolumn,showpacs,showkeys,prc,aps,amsmathamssymb]{revtex4}

\usepackage{graphicx}
\usepackage{dcolumn}
\usepackage{bm}
\begin{document}

\title{New Calculations for Phase Space Factors Involved in Double Beta Decay}

\author{ Sabin Stoica}
\email{stoica@theory.nipne.ro}
\author{Mihail Mirea}

\affiliation{Horia Hulubei Foundation, P.O. MG12, 077125-Magurele, Romania and
\\Horia Hulubei National Institute of Physics and Nuclear Engineering, P.O. Box MG6, 077125-Magurele, Romania}

\pacs{23.40.Bw, 21.60.Cs, 23.40.-s, 14.60.Pq}

\keywords{Double beta decay, phase space factors, electron wave functions}

\begin{abstract}

\noindent
 We present new results for the phase space factors involved in double beta decay for $\beta^-\beta^-$ transitions to ground states and excited $0^+_1$ states, for isotopes of experimental interest. The Coulomb distortion of the electron wave functions is treated by solving numerically the Dirac equation with inclusion of the finite nuclear size and electron screening effects, and using a Coulomb potential derived from a realistic proton density distribution in the daughter nucleus. Our results are compared with other results from literature, obtained in different approximations, and possible causes that can give differences are discussed.
\end{abstract}
\maketitle

{\it Introduction}
 Within the Standard Model (SM) the double beta decay (DBD) can occur through several decay modes, but the only measured at present is that with the emission of two electrons and two antineutrinos ($2\nu\beta^-\beta^-$) and which conserves the lepton number. However, beyond SM theories allows this process to occur without emission of neutrinos as well, and this possibility makes DBD a nuclear process of major interest for testing the lepton number conservation (LNC) and for understanding the neutrino properties. There are recent excellent reviews, containing also a comprehensive list of references \cite{AEE08} - \cite{VES12}, where the reader can find a complete information on this subject. The DBD lifetimes can be factorized, in a good approximation, as follows:

\begin{equation}
 \left( T^{2\nu}_{1/2} \right)^{-1}=G^{2\nu}(E_0, Z)\mid M^{2\nu}\mid^2,
\end{equation}

\begin{equation}
\left( T^{0\nu}_{1/2} \right)^{-1}=G^{0\nu}(E_0, Z)\mid M^{0\nu}\mid^2 \left(< \eta_{l}> \right)^2 \ ,
\end{equation}

\noindent
where $<\eta_l>$ is a beyond SM parameter containing information about the properties of the virtual particles involved in the decay within a specific mechanism, $M^{(2\nu, 0\nu)}$ are the nuclear matrix elements (NME) and $G^{(2\nu, 0\nu)}$ are phase space factors (PSF) for the corresponding decay modes(see e.g. Ref. \cite{VES12}). As seen, they are key quantities for estimating the lifetimes and/or for deriving the $<\eta_l>$ parameter, so it is very important to calculate them precisely. So far much effort has been paid for the accurate calculation of the NME. Several methods have been developed for that, the most used being: ORPA-based \cite{ROD07}-\cite{FAN11}, \cite{KOR07}-\cite{SK01}, Shell model-based \cite{CAU95}-\cite{HS10}, IBA-2 \cite{IBA2}, Energy Density Functional Method \cite{RMP10}, PHFB \cite{RAH10}. The NME have been calculated for all the transitions, decay modes and isotopes of interest, and the uncertainties in their estimation have been largely discussed during time in literature. The PSF have been calculated since long time \cite{PR59}-\cite{SC98} but they were less discussed, being considered to be computed with enough precision. Recently, they were recalculated within an improved approach by using exact electron Dirac wave functions (w.f.), taking into account the finite nuclear size and electron screening effects \cite{KI12}. The authors found differences between their results and those calculated previously with approximative electron w.f., especially for heavier nuclei. However, besides the NME, it is very important to have values of the PSF precisely calculated as well, both to improve the DBD lifetimes predictions and to extract nuclear model parameters. One example is the extraction of the $g_{pp}$ parameter in the QRPA calculations of the NME involved in DBD. In this work we report new results for the PSF involved in $2\nu$- and $0\nu$-$\beta^-\beta^-$ decay modes for transitions to the ground states (g.s.) and excited $0^+_1$ states. We developed routines for computing the relativistic (Dirac) electron w.f. taking into account the nuclear finite size and screening effects. In addition to the previous calculations, we use a Coulomb potential derived from a realistic proton density distribution in the daughter nucleus. We compare our results with other results from literature, obtained in different approximations, and discuss the causes that can give differences between different calculations.

{\it Formalism}. The PSF have been calculated first in Refs. \cite{PR59}-\cite{HS84} by using a non-relativistic approach. The distortion of the w.f. by the Coulomb field of the daughter nucleus was considered through Fermi (Coulomb) factors obtained by taking the square of the ratio of the Schr\"odinger scattering solution for a point charge $Z$ to a plane wave, evaluated at the origin. The use of such a simple expression for the Fermi factors allows us to get analytical formula for the PSF. In a better approximation, the Fermi factor is defined as the square of the ratio of the values of the Dirac $s$ w.f. of the electron at the nuclear surface $R_A= 1.2A^{1/3}$ fm \cite{SC98}:

\begin{equation}
{\it F_0(Z,\epsilon)} = 4 (2pR_A)^{2(\gamma_1 - 1)} \frac{|\Gamma(\gamma_1 + iy)|^2 \exp(\pi y)}{\left[\Gamma(2\gamma_1+1)\right]^2}
\label{eq:}
\end{equation}
\noindent
where $y = \pm\alpha Z \epsilon/p$, $\epsilon$ and $p = |{\bf p}|$ are the energy and electron momentum  and $\gamma_1 = \left[1-(\alpha Z)^2\right]^{1/2}$ with $\alpha = 1/137$. In a more rigorous treatment the electron relativistic w.f. are expressed as a superposition of $s$ and $p$ Coulomb distorted spherical waves. Their radial parts are obtained as solutions of the Dirac equations with a central field \cite{TOM91}, \cite{KI12}:
\begin{eqnarray}
{dg_{\kappa}(\epsilon,r)\over dr}=
-{\kappa\over r}g_{\kappa}(\epsilon,r)
+{\epsilon-V+m_ec^2\over c\hbar}f_{\kappa}(\epsilon,r)\\
{df_{\kappa}(\epsilon,r)\over dr}=
-{\epsilon-V-m_ec^2\over c\hbar}g_{\kappa}(\epsilon,r)
+{\kappa\over r}f_{\kappa}(\epsilon,r)
\end{eqnarray}
that depends on the relativistic quantum number
$\kappa=(l-j)(2j+1)$. The quantities
$g_{\kappa}(\epsilon,r)$ and $f_{\kappa}(\epsilon,r)$
are the small and large components of the solutions that
have the following asymptotic behavior:
\begin{eqnarray}
\left(\begin{array}{l}g_{k}(\epsilon,r)\\f_{k}(\epsilon,r)\end{array}\right)\sim\\ \nonumber
{\hbar e^{-i\delta_k}\over pr}
\left(\begin{array}{l}\sqrt{{\epsilon+m_ec^2\over 2\epsilon}}
\sin(kr-l{\pi\over 2}-\eta\ln (2kr)+\delta_k)\\
\sqrt{{\epsilon-m_ec^2\over 2\epsilon}}
\cos(kr-l{\pi\over 2}-\eta\ln (2kr)+\delta_k)\end{array}\right)
\end{eqnarray}
Here,
$c$ is the velocity of the light,  $m_e/\epsilon$ are the electron mass/energy,
$k={p/\hbar}$ is the electron wave number, $\eta=Ze^2/\hbar v$ is the Sommerfeld parameter,
$\delta_\kappa$ is the phase shift and $V$ is the Coulomb potential between the
electron and the daughter nucleus. The nuclear size corrections are usually taken into account by considering an unscreened potential $V$ obtained for a uniform charge
distribution in a sphere of radius $R_A$ \cite{DOI83}, \cite{KI12}:
\begin{equation}
V(r)=\left\{\begin{array}{ll}-{Z{\alpha\hbar c}\over r}, & r \ge R_A,\\
-Z(\alpha\hbar c)\left({3-(r/R_A)^2\over 2R}\right), & r<R_A, \end{array}
\right.
\label{cd}
\end{equation}

A further improvement in calculation is to take into account the screening effect. This can be done by multiplying the above expression of $V(r)$ with a function $\phi(r)$, which is the solution of the
Thomas Fermi equation:
$d^2\phi/dx^2 = \phi^{3/2}/\sqrt x$, with $x=r/b$, $b\approx 0.8853a_0Z^{-1/3}$ and $a_0$ = Bohr radius. It is
calculated within the Majorana method \cite{sal}. This last approach was used in Ref. \cite{KI12} to calculate the PSF.

In this work, we go further in accuracy and take into account the influence of the nuclear structure by deriving the potential $V(r)$ from a realistic proton density distribution
in the daughter nucleus. This is done by solving the Schr\"odinger equation for a Woods-Saxon (WS) potential. In this case,
\begin{equation}
V(r)=\alpha\hbar c\int{\rho_e(\vec{r'})\over \mid \vec{r}-\vec{r'}\mid}
d\vec{r'}
\label{vpot}
\end{equation}
where the charge density is
\begin{equation}
\rho_e({\vec{r}})=2\sum_i v_i^2 \mid\Psi_i(\vec{r})\mid^2
\end{equation}
$\Psi_i$ is the proton (WS) w. f.  of the single particle state
$i$ and $v_i$ is its occupation amplitude. The factor 2 reflects the
time reversal degeneracy. The screening effect is taken into account in the same manner as in Ref. \cite{KI12}.

To compute the PSF, the electron phase factors $f_{jk}^{(0)}$
must be obtained from  the solutions of the Dirac equation by neglecting the neutrino mass:
\begin{eqnarray}
f_{11}^{(0)}=\mid f^{-1-1}\mid^2+\mid f_{11}\mid^2+\mid f^{-1}_{~~~1}\mid^2
+\mid f^{~-1}_1\mid^2
\end{eqnarray}
with
\begin{eqnarray}
f^{-1-1}=g_{-1}(\epsilon_1)g_{-1}(\epsilon_2)~;~f_{11}=f_1(\epsilon_1)f_1(\epsilon_2),\\
f^{-1}_{~~~1}=g_{-1}(\epsilon_1)f_1(\epsilon_2)~;~ f_{1}^{~-1}=f_1(\epsilon_1)g_1(\epsilon_2)
\end{eqnarray}
\noindent
The values of the $f$ and $g$ functions are approximated with the solutions on the surface (the method I from \cite{KI12}).
\begin{eqnarray}
g_{-1}(\epsilon)=g_{-1}(\epsilon,R)~;~ f_1(\epsilon)=f_1(\epsilon,R)
\end{eqnarray}
\noindent
For the two neutrino $\beta\beta$ decay, the PSF are:
\begin{eqnarray}
\label{int2}
G_{2\nu}={2\tilde{A}^2\over 3\ln 2 g_A^4(m_ec^2)^2}
\int_{m_ec^2}^{T_0-m_ec^2} \int_{m_ec^2}^{T_0-\epsilon_1} \int_{0}^{T_0-\epsilon_1-\epsilon_2}\\ \nonumber
\times d\epsilon_1 d\epsilon_2 d\omega_1
 f_{11}^{(0)}w_{2\nu}(\langle K_N\rangle^2+\langle L_N\rangle^2+
\langle K_N\rangle \langle L_N\rangle)
\end{eqnarray}

\noindent
where $T_0 = Q_{\beta\beta}+2m_0c$ is the total energy released in the decay and $\langle K_N\rangle$, $\langle L_N\rangle$ are expressions (known in the theory of DBD) that depend on the electron and neutrino ($\omega_{1,2}$) energies, and on the g.s. energies of the initial nucleus and of excited states of the intermediate nucleus \cite{HS84}-\cite{KI12}. $\tilde{A}=1.12A^{1/2}$ (in MeV) gives the energy of the giant Gamow-Teller resonance in the intermediate nucleus, and
\begin{equation}
w_{2\nu}={g_A^4(G\cos\theta_C)^4\over 64\pi^7\hbar}
w_1^2w_2^2(p_1c)(p_2c)\epsilon_1\epsilon_2.
\end{equation}
\noindent
The PSF are finally renormalized to the electron rest energy and are reported in $[yr^{-1}]$.

For the $0\nu\beta\beta$ decay, the PSFs are
\begin{equation}
\label{int1}
G_{0\nu}={2\over{4g^4_AR^2 \ln2}}\int_{m_ec^2}^{T_0-m_ec^2}
f_{11}^{(0)}w_{0\nu}d\epsilon_1
\end{equation}

\begin{table*}
\caption{Values of the PSF $G_{2\nu} [yr^{-1}]$ for transitions to g.s. (first line) and to excited $0_1$ state(second line). The relative differences in percentage $\varepsilon[\%]$ between other previous calculations (indicated by references) and our results are displayed in the last columns. The $Q_{\beta\beta}$ and $Q_{\beta\beta}^1$ are the kinetic energies available in the corresponding decays}.
\begin{tabular}{|c|c|c|c|c|c|c|c|c|c|c|c|c|}
\hline

\multicolumn{13}{|c|}{ $G_{2\nu} [yr^{-1}]$ } \\ \hline
Nucleus                   &$^{48}$Ca&$^{76}$Ge&$^{82}$Se& $^{96}$Zr&$^{100}$Mo&$^{110}$Pd&$^{116}$Cd&$^{128}$Te&$^{130}$Te&$^{136}$Xe&$^{150}$Nd&$^{238}$U\\ \hline
$Q_{\beta\beta}$          & 4.272   & 2.039   &   2.995 &   3.350  &  3.034   &   2.018  &  2.814   &  0.866   &   2.527  &    2.458 &  3.371   & 1.145 \\
$Q_{\beta\beta}^1$        & 1.275   & 0.917   & 1.507   &   2.202  &    1.904 &  0.548   &   1.057  &          &   0.733  &    0.879 &    2.630 & 0.204 \\ \hline
\cite{DOI85}-\cite{DOI93} &1.62E-17 &5.38E-20 &1.83E-18 &          &3.86E-18&          &          & 3.50E-22 & 1.97E-18 & 2.03E-18 & 4.87E-17 &       \\ \hline
\cite{SC98}               &1.62E-17 & 5.26E-20&1.74E-18 & 7.28E-18 & 3.60E-18 &          & 2.99E-18 & 3.44E-22 & 1.94E-18 & 1.98E-18 & 4.85E-17 &       \\
                          &3.76E-22 & 7.69E-23&4.80E-21 & 1.90E-19 &1.01E-19  &          &8.90E-22  &          & 1.86E-20 & 4.85E-22 & 4.85E-18 &     \\ \hline
\cite{KI12}               &1.56E-17 &4.82E-20 &1.60E-18 & 6.82E-18 & 3.31E-18 & 1.38E-19 & 2.76E-18 & 2.69E-22 &1.53E-18  &1.43E-18  & 3.64E-17 & 1.46E-20\\
                          &3.63E-22 &6.98E-23 &         & 1.75E-19 & 6.06E-20 & 4.84E-24 & 8.73E-22 &          &7.57E-23  &3.62E-22  & 4.33E-18 & 4.64E-25\\ \hline
Present 		              &1.55E-17 & 4.39E-20&1.48E-18 &5.94E-18  &2.91E-18  &1.20E-19  &2.58E-18  &2.53E-22  &1.46E-18  &1.37E-18  &3.42E-17  & 1.15E-19\\
 work                     &3.70E-22 & 5.95E-23&3.71E-21 &1.31E-19  &4.58E-20  &3.32E-24  &6.95E-22  &          &7.71E-23  &3.56E-22  &3.99E-18  & 3.26E-24\\ \hline
$\varepsilon$ \cite{DOI85}-\cite{DOI93}&4.3&18.4&-2.2&&4.1&&&27.7&25.9&32.5&29.8& \\ \hline	
$\varepsilon$ \cite{SC98}	 & 4.3  & 16.5 & -7.5 & -4.4 & -2.8 &   & 13.7 & 26.5 & 24.7 & 30.8 & 29.5 &   \\
                           & 1.6 & 22.6 & 22.7 & 31.1& 54.7 &   & 21.9 &   & 99.6 & 26.6 & 17.7 &\\ \hline
$\varepsilon$ \cite{KI12} & 0.6 & 8.9 & 7.5 & 12.9 & 12.1 & 13.0 & 6.5 & 5.9 & 4.6 & 4.2 & 6.0 & $>$100 \\
                          & -1.9 & 14.8 &   & 25.1 & 24.4 & 31.4 & 20.4 &   & -1.8 & 1.7 & 7.9 & $>$100\\ \hline

\multicolumn{13}{|c|}{ $G_{0\nu} [yr^{-1}]$ } \\ \hline
\cite{DOI85}-\cite{DOI93} &2.61E-14&2.62E-15&1.14E-14& &1.87E-14& & &7.48E-16&1.94E-14&1.94E-14 &8.59E-14&\\ \hline	
\cite{SC98}               &2.60E-14&2.55E-15&1.11E-14 &2.31E-14&4.56E-14& &1.89E-14&6.71E-16&1.67E-14&1.77E-14&7.84E-14&\\ \hline	
\cite{KI12}               &2.48E-14&2.36E-15&1.02E-14&2.06E-14&1.59E-14&4.82E-15&1.67E-14&5.88E-16&1.42E-14&1.49E-14&6.30E-14&3.36E-14\\
                          &2.99E-16&1.78E-16&        &4.57E-15&3.16E-15&8.84E-17&7.16E-16&        &3.09E-16&6.13E-16&2.73E-14&7.53E-16\\ \hline	
Present 		              &2.49E-14&2.34E-15&1.01E-14&2.03E-14&1.57E-14&4.79E-15&1.66E-14&5.55E-16&1.41E-14&1.46E-14&6.20E-14&3.11E-14\\
work                      &3.05E-16&1.87E-16&9.17E-16&3.30E-15&3.07E-15&1.08E-16&7.19E-16&        &3.57E-16&6.59E-16&2.70E-14&1.18E-15\\ \hline	
$\varepsilon$ \cite{DOI85}-\cite{DOI93} & 4.6&10.7&11.4& &16.0 & & &25.8&27.3&24.7&27.8& \\ \hline					
$\varepsilon$ \cite{SC98}	  & 4.2  & 8.2  & 9.0  & 12.1  & 65.6  &  & 12.2  & 17.3  & 18.4  & 17.5  & 20.9  & \\ \hline
$\varepsilon$ \cite{KI12}  & -0.4  & 0.9  & 1.0  & 1.5  & 1.3  & 0.6  & 0.6  & 5.6  & 0.7  & 2.0  & 1.6  & 7.4 \\
                          &-2.0  & -5.1  &  & 27.8  & 2.9  & -22.2  & -0.4  &  & -15.5  & -7.5  & 1.1  & -57 \\ \hline	
\end{tabular}
\label{tabelx1}
\end{table*}

where
\begin{equation}
w_{0\nu }={g_A^4(G\cos\theta_C)^4\over 16\pi^5}
(m_ec^2)^2 (\hbar c^2) (p_1c) (p_2c) \epsilon_1\epsilon_2
\end{equation}
\noindent
where $G=1.16637\times10^{-5}$ GeV$^{-2}$ is the Fermi constant and $\cos\theta_C$=0.9737 \cite{TOM91}.
In Eq.(16) it is convenient to redefine the PSF by a renormalization that eliminates the constant $g_A$ and correlates (by dividing by $4R_A^2$) the dimension of $G_{0\nu}$ with the NME which are dimensionless. Thus, our PSF are reported in $[yr^{-1}]$.

{\it Results}. The single particle densities inside the daughter nucleus, needed to derive the potential $V(r)$, are obtained by solving the Schr\"odinger equation for a spherical WS potential, including spin-orbit
and Coulomb terms. The universal parametrization was employed as
in Ref. \cite{mir}. The occupation amplitudes are obtained within the BCS
approach \cite{hill}. Further, the Dirac equation is solved for the electron moving
in the potential $V(r)$, created by the proton distribution, by using the
power series method from Ref. \cite{buh}. We built up a numerical code that use an algorithm similar to that used in Ref. \cite{sal1}. The asymptotic normalization to unity is done as in Ref. \cite{sal2}.
The solutions of the electron w. f. are computed numerically by approximating them
 with infinite polynomials whose coefficients at different distances $r$ are connected
analytically by the particular forms of the Dirac equations
and by the values of the Coulomb potential. Therefore, the numerical values of the w.f. can be
calculated step by step, by increasing the distance $r$.
 At very large distances, the behavior of the w. f.
 must resemble to that of the Coulomb function. This last
condition provides a way to renormalize the amplitude to unity
 and to determine the phase shift.
To solve the integrals (14) and (16), we compute the values of the electron w. f. and these values are interpolated. 
Because the
the w. f. values at the nuclear surface vary rapidly for energies close
to $m_0c^2$, we took additional mesh points in the vicinity of this region for improving the numerical accuracy. Our results are presented in Tables I for twelve nuclei of experimental interest. In the first (upper) part of the table the PSF values, $G_{2\nu}$, for the $2\nu\beta\beta$ decay mode, for the transitions to the g.s (first row in the box) and to excited $0^+_1$ states (second row in the box) are displayed. For comparison, similar results are also displayed, indicating the references where they are taken from. The maximum available kinetic energies $Q_{\beta\beta}$ and $Q_{\beta\beta}^1$ for the transitions to the g.s. and excited $0^+_1$ states, respectively, are given as well. In the last rows the relative differences in percentage between other results and ours \{[$\epsilon = G_{2\nu}(ref.)- G_{2\nu}(our)]/G_{2\nu}(ref.)$\} are showed for comparison. The PSF values, $G_{0\nu}$, for the $0\nu\beta\beta$ decay mode and the relative differences are presented in a similar way in the second (lower) part of the table. The relative differences between our $G_{2\nu}$ values and other results for the transitions to g.s. are within 18.4\%, for the light nuclei. For the heavier nuclei, with $A>128$, where the influence of the potential V(r) is stronger, the relative differences are larger (between [25 - 32.5]\%), as compared with Refs. \cite{DOI85}, \cite{SC98}, while the agreement with Ref. \cite{KI12} is excellent with one exception: $^{238}U$.  For the transitions to the excited $0^+_1$ state the agreement with previous results is within 31.4\%, with a few exceptions ($^{100}Mo$ and $^{130}Te$ from \cite{SC98}, and $^{238}U$ from \cite{KI12}), which should be revised. For $G_{0\nu}$ the agreement between our results and the previous ones is better than in the $2\nu$ case, and follows the same features discussed above. We notice the excellent agreement with the results from Ref. \cite{KI12} for the transitions to g.s., but, also, notice some few cases that should be revised($^{100}Mo$, and $^{238}U$ for transitions to $0_1^+$ state). The differences in PSF values could mainly come from two sources: the quality of the approach and the accuracy of the numerical methods, that are used. On the one hand it is clear that an improved  treatment of the electron w.f. (relativistic treatment with inclusion of finite nuclear size and electron screening effects, and using a realistic Coulomb potential) is preferable to a less rigorous one, as it is also highlighted in \cite{KI12}. Related to this, since the influence of the structure of the daughter nucleus was never investigated, we performed the calculations with expressions of V(r) given either by a uniform charge distribution or derived from a realistic proton density distribution. The differences, we got between the two calculations are within $5\%$ for both $G_{0\nu,2\nu}$. On the other hand, as we already mentioned, the $f$ and $g$ functions take their maximum values in the vicinity of $m_0c^2$ and, hence, it is necessary a more rigorous treatment of the numerical integration in that region. An inadequate numerical treatment can change significantly the results.

{\it Conclusions} In summary, we performed an independent calculation of the PSF involved in $\beta^-\beta^-$ decays modes, for transitions to the g. s. and excited $0^+_1$ states, for twelve nuclei of experimental interest. The Coulomb distortion of the electron w. f. is obtained by solving numerically the Dirac equation including the finite nuclear size and electron screening effects. In addition to other previous approaches, we used a Coulomb potential derived from a realistic proton density distribution in the daughter nucleus. The relative differences between other results and ours are within $\approx$ 32\%, with a few exceptions that have to be revised. The differences between the PSF values can come, in part, from the rigor of the approach used in their calculation and, in part, from the accuracy of the numerical method used for integration. Since the PSF are important ingredients both for the estimation of the  DBD lifetimes and for the extraction of some key nuclear model parameters, a deeper investigation of these issues and a rigorous calculation of them is still very needed.

{\it Acknowledgments}This work was supported by a grant of the Romanian Ministry of National Education, CNCS UEFISCDI, project PCE-2011-3-0318, Contract no. 58/28.10/2011.

\end{document}